\documentclass{acmconf}
\usepackage{epsfig}
\usepackage{amsmath, amsthm, amssymb}

\begin{document}
\title{A Game Theoretical Approach to Broadcast Information Diffusion in Social Networks}
\author{D.~Zinoviev and V.~Duong\\%
Mathematics and Computer Science Department, Suffolk University\\%
Boston, Massachusetts 02114, USA%
}
\maketitle

\vskip\baselineskip

\abstract{One major function of social networks (e.g., massive online social networks) is the dissemination of information, such as scientific knowledge, news, and rumors. Information can be propagated by the users of the network via natural connections in written, oral or electronic form. The information passing from a sender to receivers and back (in the form of comments) involves all of the actors considering their knowledge, trust, and popularity, which shape their publishing and commenting strategies. To understand such human aspects of the information dissemination, we propose a game theoretical model of a one-way information forwarding and feedback mechanism in a star-shaped social network that takes into account the personalities of the communicating actors.}

\section{INTRODUCTION AND PRIOR WORK}

A social network is a collection of actors (network members) connected with links---indicators of proximity. One major function of social networks is information and innovation diffusion. Information represents scientific knowledge, news, rumors, etc.~\cite{1099206,nekovee2008,watts2002,zanette2002}. As an important form of social organization, information can shape public opinion, inform and misinform the society, cause panic in a society, promote products, etc.~\cite{nekovee2008}. It is disseminated by the members of the network via natural connections in written, oral or electronic form. 

Due to its importance, information diffusion has been one of the focuses in social network research. For example, theories of rumor spreading are proposed by Zanette~\cite{zanette2002} and Nekovee {\it et al.}~\cite{nekovee2008}. Game theoretical approach to information propagation (namely, to learning) has been suggested by Gale \& Kariv~\cite{gale2003} and Acemoglu {\it et al.}~\cite{NBERw14040}. Ellwardt \& van Duijn explored gossiping in small organizational social networks~\cite{ellwardt2009}. Since information dissemination and other various social network activities are supported by the structural organization of social networks, social network topology receives a lot of research attention. An effect of network topology on the information diffusion was observed by Hirshman \& Carley~\cite{hirshman2009}: sparse networks are more effective for information entrance, and clustered (cellular) network structure decreases information diffusion.

In this paper, we propose a game theoretical model for the one-way information passing from a selected member of a social network (the sender, the speaker) to his immediate network neighborhood (the friend list, the ``first circle,'' the receivers or the listeners). The novelty of our model is that psychological characteristics are explicitly modeled in information dissemination. In our model, information passing intrinsically involves all parties considering their psychological characteristics: self-perceived knowledge, trust, and popularity---which further determine their decisions of whether or not to forward the information and whether or not to provide feedback (comments).

Feedback in information dissemination is explicitly considered as strategic moves in our game theoretical model. Related to our work, Lampe {\it et al.}~\cite{1099206} also analyzed the mechanism of feedback, its influence on the members of online communities, and its role in learning transfer. Similar concept of social influence, but in the context of community building, has been researched by Crandall {\it et al.}~\cite{crandall2008}.

This paper is an extension of the results described by Zinoviev {\it et al.}~\cite{zinoviev2010a,zinoviev2010}. In the first original paper~\cite{zinoviev2010}, we presented an analysis of a one-directional atomic peer-to-peer communication, where one actor is a speaker (he sends a message) and the other is a listener (and optionally a commenter---she receives the message and sends a reply), and the comment, if any, is indivisible from the original message. In the second paper~\cite{zinoviev2010a}, we removed the one-way limitation and allowed two actors to communicate in a full duplex mode, when either or both actors are speaking, listening, and commenting at the same time.

In this paper, the two-way peer-to-peer model is replaced by a one-way atomic broadcast (``friendcast''), which is a more realistic scenario in massive online social networks. Examples of broadcast include wall posts in Facebook with follow-up comments, blog posts (with comments) in LiveJournal and Blogger, and tweets in Twitter. Additionally, we reorganized the knowledge model by introducing knowledge forgetfulness, willingness to learn, and fuzzy measures of knowledge and belief. We also present some early experimental evidence of the correlation between a speaker's posting rate and his popularity among the listeners.

The rest of the paper is organized as follows. Section~\ref{overview} presents the overview of the model. Section~\ref{transfer} introduces the knowledge and feedback transfer mechanisms. Section~\ref{gamematrix} describes the organization of the game matrix. Some simulation results illustrating the model are presented in Section~\ref{simulation}. Finally, section~\ref{conclusion} outlines the key results and future research directions.

\section{MODEL OVERVIEW\label{overview}}
The proposed model consists of five major components: the network model, the knowledge model, the popularity model, the trust model, and the utility definition. Once the utility is defined for all actors, standard game theoretical methods are applied to calculate the optimal strategies for the actors.

\subsection{Network Configuration}

\begin{figure}[b!]\centering
\epsfig{file=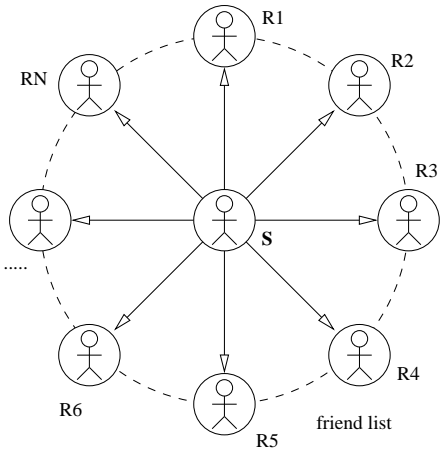,width=0.7\columnwidth}
\caption{\label{network}Communication between the sender $S$ and the receivers $R_i$.}
\end{figure}

We consider a star-shaped network with one hub (the sender or the speaker, $S$) connected to a collection of $N$ independent receivers (listeners, $R_i$). The connections are bidirectional: new information propagates from the sender to the receivers, and feedback (if any) propagates from the receivers to the sender (Figure~\ref{network}).

If the sender decides to share information with the neighbors, then the information is shared with all N receivers at once. Individual receivers may choose to respond to the broadcast message or not, depending on their optimal strategies. We assume that if the receivers respond, they do so in the order of their position in the sender's friend list (the first friend responds first, followed by the second one, etc.). No further messages are sent by S until all receivers respond or indicate their unwillingness to respond. The composition of the sender's friend list does not change while the message exchange is in progress. In this sense, the information diffusion model described in the paper is an atomic broadcast (``friendcast''). 

\subsection{Knowledge Model}

We represent knowledge as a finite collection of enumerated elementary assertions $\mathbf{A}=\{A_i\}$ (a knowledge base). For example, $A_0=$ ``the world is flat'' is an assertion, and so is $A_1=$ ``the world is round.'' The meaning of the word ``elementary'' is application-specific. The model can operate at the level of data (facts), information (processed facts) or knowledge (patterns)~\cite{tuomi99}. Each individual in the system knows from 0 to $|A|$ assertions from the collection.

In contrast to our previous model~\cite{zinoviev2010a,zinoviev2010}, an assertion in the new model is a two-component tuple $A_i=\{k_i, b_i\}$, $0\le k_i\le 1$, $-1\le b_i \le 1$. The first element of the tuple, $k_i$, is the quantitative measure of knowledge. The quantity of 1 represents full knowledge of the assertion. The quantity of 0 represents no knowledge. A value of $0<k_i<1$ corresponds to partial knowledge (such as ``I know some quantum physics'').

The second element of the tuple, $b_i$, is the self-perceived quality of the assertion, or the belief associated with the assertion. The belief of 1 represents a true self-perceived fact (``I believe that the world is round''); the belief of -1 represents a false fact (``I do not believe that the world is flat''); the belief of 0 represents an assertion whose correctness cannot be established, or a rumor (``There may be no life on Alpha Centauri'').

As an example, assertion ``I know some astrology, but I don't really believe in it'' may look like $A_i=\{0.3, -0.9\}$.

The value of an assertion is the product of the knowledge and the belief: $a_i=k_i\times b_i$. We propose that the average knowledge $K$ of an actor characterizes the actor's ability to apply reasoning to unknown assertions:
\begin{equation}
K=\overline{|a_i|},\;0\le K \le 1.
\end{equation}

\subsubsection*{Ontology}
Some assertions in the knowledge base may be interdependent (correlated or anti-correlated). The $|A|\times |A|$ matrix $\mathbf{M}$, $-1\le m_{ij} \le 1, m_{ii}=1$, represents the correlations between assertions. We call $M$ an ontology matrix. $m_{ij}$ is the correlation coefficient between $A_i$ and $A_j$: the change of belief in $A_i$ may cause a change of belief in $A_j$, $\Delta b_j=m_{ij}\Delta b_i$. For example, $m_{01}=m_{10}=-1$ for the ``world'' assertions mentioned above. In general, $m_{ij}\ne m_{ji}$. 

\subsubsection*{Knowledge Dynamics}

Two new elements of the current model are an individual's forgetfulness and willingness to learn.

We propose that knowledge of an individual deteriorates over time. At each time step, 
\begin{equation}
k_{i,t+1}=\sqrt{\zeta}\; k_{i,t};\;
b_{i,t+1}=\sqrt{\zeta}\; b_{i,t};\;
i=1\ldots |A|,\label{zeta}
\end{equation}
where $0\le\zeta\le 1$ is a measure of remembrance. $\zeta=1$ represents no forgetfulness (no knowledge deterioration); $\zeta=0$ represents complete forgetfulness.

We propose that an individual receiver may be resistant to learning from the sender. The willingness to learn is controlled by yet another coefficient $0\le w\le 1$: $w=0$ represents a ``stubborn'' individual who is not willing to learn; $w=1$ represents an enthusiastic knowledge adopter.

\subsubsection*{Operations on Assertions}

In the process of learning, instances of assertions are combined using the learning operator $\oplus$, which is defined below.

The operator affects assertion tuple components in different ways. The functionality of the operator satisfies the following axioms:
\begin{enumerate}
\item For each assertion, there exists an instance of the assertion with full knowledge, $I_i=I_i\{1,\bullet\}$;
\item For each assertion, there exists an instance of the assertion with no knowledge, $O_i=O_i\{0,\bullet\}$;
\item $A_i\{k,\bullet\}\oplus O_i=A_i\{k,\bullet\}$ (learning an instance of an assertion with no knowledge does not increase knowledge);
\item $A_i\{k,\bullet\}\oplus I_i=I_i$ (an instance of an assertion with full knowledge absorbs any other instance of the assertion);
\item $A_i\{k_x,\bullet\}\oplus A_i\{k_y,\bullet\}\le I_i$ (two instances of the same assertions, when combined, do not create more knowledge than it is possible);
\label{lastaxiom}
\end{enumerate}

The knowledge contained in two partial instances of the same assertion $A_i$  ($k<1$) may or may not intersect ($A_x=$``I know some quantum theory'' and $A_y=$``I know some quantum theory'' do not necessarily refer to the same ``some.'') Therefore, combining them in the case of full overlap produces $\max\left(k_x,k_y\right)$ knowledge and in the case of least overlap---$\min\left(1, k_x+k_y\right)$ knowledge. Assuming that both cases are equally probable, the average combined knowledge is given by the following equation:
\begin{equation}
k_{x\oplus y}=\frac{\max\left(k_x,k_y\right)+\min\left(1,k_x+k_y\right)}{2}.
\end{equation}

For the sake of simplicity, this piecewise linear function can be approximated by a smooth function:
\begin{equation}
k_{x\oplus y}=k_x+k_y-k_xk_y=k_x+k_y\left(1-k_x\right).\label{combine}
\end{equation}

A similar equation can be written for the new belief component, except that the added belief is weighted by the corresponding knowledge to avoid having strong influence by confident but not knowledgeable actors:
\begin{equation}
b_{x\oplus y}=b_x+k_yb_y\left(1\pm b_x\right).
\end{equation}
The sign in the parentheses depends on whether the added belief is positive (``--'') or negative (``+'').

Here is the complete definition of the learning operator for $k_x+k_y>0$:
\begin{equation}
A_{ix}\oplus A_{iy}=\left\{k_x+k_y\left(1-k_x\right),b_x+k_yb_y\left(1\pm b_x\right)\right\}.\label{oplus}
\end{equation}
One can prove that this definition satisfies Axioms 1--\ref{lastaxiom}.

\subsection{Popularity Model}

The general popularity framework has been described by Parkhurst \& Hopmeyer~\cite{parkhurst98} and Coie {\it et al.}~\cite{Coie82}. They distinguish two types of popularity: sociometric popularity $P_S$ and peer-perceived popularity $P_P$. The sociometric popularity of an actor is calculated in terms of the number of the actor's friends $F\!F$ and enemies $EE$. Sociometric popularity is often expressed in terms of social preference $SP=F\!F-EE$ and social impact $SI=F\!F+EE$.

The perceived popularity is a measure of the actor's social dominance or influence and is one of the components of his utility. Perceived popularity can be high ($P_P=1$), average ($P_P=0$) or low ($P_P=-1$).

It follows from Table 1 in~\cite{parkhurst98} that these two measures of popularity are correlated:
\begin{equation}
P\equiv P_P\sim\left(F\!F/2-EE/10\right)\sim F\!F.\label{eq:popularity}
\end{equation}
In other words, the perceived popularity of an actor is proportional to the number of the actor's friends (unless he has a lot of enemies). The value of $F\!F$ is easily computed (it is the degree of a node representing the actor in a social network graph) and can be used to estimate $P$.

In our model, we assume that a speaker or commenter improves his popularity by posting assertions or comments. Otherwise, the popularity gradually declines over time. This means that the actors with a high publishing rate (the number of publications per unit time) will see a steady growth of $P$ and, according to Eq.~(\ref{eq:popularity}), the corresponding growth of the number of listeners. On the other hand, idle (silent) actors will become less popular and will slowly lose listeners.

It is not clear yet if actors' publishing rates and audience growth rates in real social networks are indeed linearly correlated. However, we have some preliminary evidence based on experiments with LiveJournal (a blogging Web site) that this assumption may be at least partially true.

\subsection{Trust Model}

We define actor-to-actor trust $0\le T^{XY}\!\le 1$ as a measure of the extent to which actor $X$ trusts the information provided by another actor $Y$. If $T^{XY}=0$, then $X$ disregards $Y$'s publications and comments. If $T^{XY}=1$, then $X$ unconditionally accepts $Y$'s publications and comments. In general, $T^{XY}\ne T^{YX}$ and $T^{XX}=1$.

Each actor $X$ has reputation $R_X=\overline{T^{Y\!X}}, X\ne Y$, which is the average trust of all other actors in $X$. As we will show in the next section, $R_X$ is one of the components of X's utility.

\section{KNOWLEDGE AND FEEDBACK TRANSFER\label{transfer}}
In this section, we describe the knowledge transfer process and the associated utility function.
The transfer between the sender and receivers involves learning, feedback, and utility updating of all participating actors.

\subsection{Utility and Personality}

A single communication between the sender and the receivers consists of an exchange of at most one assertion (from $S$ to $R$s) and up to $N$ feedback messages. The communication intrinsically involves both actors considering their self-perceived knowledge, reputation, and popularity, which further determine their decisions of whether or not to publish the assertion in the first place and whether or not to provide feedback. An actor's self-perceived knowledge, reputation, and popularity collectively form her utility. The weights an actor puts on these three utility components characterize this actor's personality.

We believe that the purpose of a rational actor $X$ is to maximize her utility $U_X$, defined as a convex combination of knowledge, reputation, and popularity with coefficients $0\le\kappa,\rho,\pi\le1,\kappa+\rho+\pi=1$: 
\begin{equation}
 U_X=\kappa K_X+\rho R_X+\pi P_X.\label{payoff}
\end{equation}

We use a particular set of coefficients $\{\kappa,\rho,\pi\}$ to characterize a particular type of actors' personality. For example, $\kappa=\rho=0,\pi=1$ describe a network of ``Internet trolls'' (actors, for whom bloated popularity is the primary goal of networking). On the other hand, $\kappa=\rho=0.5,\pi=0$ probably corresponds to a scientific community of knowledge seeking altruists who care about their reputation and wisdom, but not about being quoted or even published.

In this paper, we focus on a homogeneous network where all actors have the same utility function coefficients. We understand that in a real social network, actors are heterogeneous. We leave the heterogeneous network as future work.

\subsection{Knowledge Transfer}

Transferring knowledge about an assertion $A_i$ from the sender $S$ to a receiver $R$ involves combining the old receiver's knowledge about this and possibly other correlated assertions $A^R_j$, forgotten to a certain extent, with the newly received knowledge $A^S_i$ (in these equations, we treat tuple $A_i$ as a two-component vector):
\begin{equation}\label{learning}
A^R_{j,(t+1)}=\left(F A^R_{j,t}\right)\oplus \left(G_{ij} A^S_{i,t}+ H_{ij}\right)\!,\;j=1\ldots |A|.
\end{equation}

Matrix $F$ describes forgetfulness and is simply another notation for Eq.~(\ref{zeta}). 
It reduces the value of each assertion (both knowledge and belief are affected). This matrix is applied to all assertions held by the sender and by all receivers at each time step, regardless of whether there was knowledge transfer or not. $F$ is defined as follows:

\begin{equation}\label{PsiMatrix}
F=\begin{bmatrix}
\sqrt{\zeta}&0\\
0&\sqrt{\zeta}
\end{bmatrix}
\end{equation}

Altogether, the left operand of the learning operator in Eq.~(\ref{learning}) describes the knowledge deterioration.

Matrix $G$ describes knowledge acceptance. It reduces the quantity of knowledge in the transferred assertion $A^S_i$ by $\left(1-w\right)$ due to the unwillingness to learn and reflects the process of calculating the actual belief in the newly acquired knowledge. The receiver accepts the amount of belief in the assertion proportional to her trust in the sender. Note that the coefficient $m_{ij}$ from the ontology matrix $M$ allows the receiver to adjust beliefs of the assertions correlated with $A^S_i$ (including, naturally, the receiver's instance of the transferred assertion, $A^R_i$). $G$ is defined as follows:

\begin{equation}\label{XiMatrix}
G_{ij}=\begin{bmatrix}
w\delta_{ij} & 0\\
0 & m_{ij}T^{RS}
\end{bmatrix}\!,\;\delta_{ij}=\begin{cases}
0&\text{if $i\ne j$},\\
1&\text{if $i=j$}.
\end{cases}
\end{equation}

The transferred assertion, as perceived by the corresponding receiver, is $G_{ij} A^S_{i,t}$.

$H$ is the vector of self-assessed knowledge. When the receiver does not have enough trust in the sender, she tries to make an educated guess about the true nature of the transferred assertion $A^S_i$, based on her own knowledge, through the ontology matrix $M$. The belief associated with $A^S_i$ is estimated as $\overline{m_{ki} b_k^R}$---the average aggregate belief of the receiver weighted by the correlation coefficients. It is further multiplied by $m_{ij}$ to reflect the potential impact of the assertion $A^S_i$ on $A^R_j$ (remember that $m_{ii}=1$). Finally, only a fraction of the estimated belief is taken into account, proportional to the lack of trust of R in S:

\begin{equation}\label{HMatrix}
H_{ij}=\left\{0,m_{ij}\left(1-T^{RS}\right)\overline{m_{ki} b_k^R}\right\}
\end{equation}

There is no knowledge quantity transfer associated with the self-assessment.

Thus, Eq.~(\ref{learning}) explains how the receiver's new knowledge is formed by combining the receiver's pre-existing (though partially deteriorated) knowledge with a mix of the learned knowledge and the receiver's educated guess.

\subsection{Trust and Popularity Calculation}

As a result of the knowledge transfer, the values of trust, reputation, and popularity of all involved actors also change.

The change of the sender's popularity is calculated as the average change of the receivers'  total knowledge caused by the acceptance of the transferred assertion:

\begin{equation}\label{DeltaP}
\Delta P^S=\overline{|a_{i,(t+1)}^R-a_{i,t}^R|}.
\end{equation}

In other words, if the receivers learn nothing new from the sender, his popularity does not change (in fact, in our model an actor's popularity decreases by the factor of $\left(1-\delta p\right)$ per time step if he or she neither publishes an assertion nor provides feedback).

The change of a receiver's trust in the sender is calculated as the difference between the sender's and the old receiver's perceptions of the transferred assertion:

\begin{equation}\label{DeltaT}
\Delta T^{RS}=1-|b_i^S-b_i^R|.
\end{equation}

In other words, agreement ($b_i^S=b_i^R$) improves trust, and disagreement ruins it. 

To make sure that neither trust nor popularity exceeds 1, we calculate the new values of $P$ and $T$ by combining the old historical values and the increments. We use different combination rules. Since popularity, according to our model, is monotonically increasing (except when the actor does not publish an assertion or feedback), we combine the old popularity and the increment the same way we combine knowledge in Eq.~\ref{combine}: 

\begin{equation}
P^S_{t+1}=P^S_t+\Delta P^S-P^S_t\Delta P^S.
\end{equation}

On the contrary, trust can both increase and decrease. Coefficient $\xi\in[0,1]$ is a measure of influence of trust history on the future trust; $\xi=0$ represents the case when the new value of trust is calculated without looking at the history; $\xi=1$ means that the new value is taken directly from the history (which makes it a constant):

\begin{equation}
T^{RS}_{t+1}=\xi T^{RS}_t+\left(1-\xi\right)\Delta T^{RS}
\end{equation}

\subsection{Feedback Transfer}
The feedback transfer from the receivers to the sender is similar to the transfer in the opposite direction, except that:
\begin{itemize}
\item the transferred assertion has no associated quantity of knowledge (only belief, because it is a response to another assertion) and 
\item the transfer does not cause the sender's knowledge deterioration (because it happens at the same time step as the original knowledge transfer).
\end{itemize}

The equations~(\ref{PsiMatrix}), (\ref{XiMatrix}), and~(\ref{HMatrix}) can be rewritten accordingly:

\begin{equation}\label{PsiMatrixFB}
F=\begin{bmatrix}
1&0\\
0&1
\end{bmatrix}
\end{equation}
\begin{equation}\label{XiMatrixFB}
G_{ij}=\begin{bmatrix}
0 & 0\\
0 & m_{ij}T^{SR}
\end{bmatrix}
\end{equation}

\begin{equation}\label{HMatrixFB}
H_{ij}=\left[0,0\right]
\end{equation}

Note that Eq.~(\ref{XiMatrixFB}) uses the trust of the sender in the receivers, $T^{SR}$, not the other way around.

The trust and popularity changes for the receiver are calculated using Eq.~(\ref{DeltaP}) and Eq.~(\ref{DeltaT}), with a suitable substitutions $S\leftrightarrow R$ and $SR\leftrightarrow RS$. When feedback is expected from more than one receiver, the increments are applied to the sender's values in the order of the receivers' occurrence on the sender's friend list.

\section{GAME MATRIX\label{gamematrix}}

In the previous section, we gave a detailed analysis of the basic steps involved in the information transmission between a sender and receivers, and described how the actors update their utilities (including knowledge, trust, and popularity) depending on whether or not assertions and feedback messages are transmitted. However, we have not answered this question yet: under what circumstances are the actors willing to transmit assertions and send feedback? In this section, we will address this question under the assumption that all actors know that each of them attempts to maximize his or her own utility, and they are fully aware of the impact on their own utilities from any combination of their individual choices.

Such a strategic interaction between the actors can be naturally modeled as a game with the actors being players. More specifically, the actors play a non-cooperative non-zero-sum $(N+1)$-person rectangular game. The utility changes of the players in the game are given by the $(N+1)$-dimensional payoff matrix $M_{sr_1r_2\ldots r_N}$ ($s,r_k\in\{1,2\}$) and depends on the strategies $s$ and $r_k$ selected by the players: the sender and the receivers. The cells of the matrix correspond to the combinations of the available actions of the actors. The sender has to choose between two actions:
\begin{enumerate}
\item[$S_1$:] not to send an assertion,
\item[$S_2$:] to send an assertion.
\end{enumerate}

Each receiver has to choose between another two actions:
\begin{enumerate}
\item[$R_1$:] not to provide feedback,
\item[$R_2$:] to provide feedback (if possible).
\end{enumerate}

Obviously, any combination of the form $\{S_1,\ldots,R_2,\ldots\}$ is infeasible: one cannot send feedback to an assertion that has never been sent in the first place.

The solution of the resulting game (the Nash equilibrium: the optimal combination of the actions, or strategies, for each player) can be found using well-known game theoretical methods (see, e.g.,~\cite{mckinsey2003}).

It can be proved that from the sender's point of view, there is always exactly one optimal strategy (to forward or not to forward)---a so-called pure strategy, regardless of the personalities of the receivers. The proof is very similar to the one presented in~\cite{zinoviev2010}. Unfortunately, in the presence of more than one receiver, it is not possible to prove that each receiver has an optimal pure strategy. Receivers may be forced to alternate the available feedback strategies to achieve a mixed Nash equilibrium. We can speculate that in real life the number of communications between the sender and the receivers is finite and that if they are substantially spaced in time, each session can be treated as independent, thus resulting in a pure-strategy game. In this case, the strategies chosen by the players are not necessarily truly optimal.

\section{\label{simulation}SIMULATION}

To observe our model's behavior, we simulated information diffusion in a loose one-shot network of actors. The network consisted of 100 unconnected members. One third of the members (``gurus'') had high average knowledge of 0.9; another third (``ignoramuses'') had low knowledge of 0.1; finally, the remaining members (``mediocres'') had medium knowledge of 0.5. The number of facts in the system was 10.

At each simulation step, a random person from the population was chosen as the sender, and another $N=1$ people were chosen as receivers. The receivers were temporarily connected to the sender. The sender and the receivers would then play the information transmission game, select the optimal strategy, and apply it to their states, thus updating the values of $k_i$, $b_i$, $a$, $T^{RS}$, $T^{SR}$, and $P$. After that, the connections between the involved actors were severed. At every 500th step, knowledge distribution in the network was calculated and plotted against the simulation time. The simulation was repeated 50,000 times.

\begin{figure}[b!]\centering
\epsfig{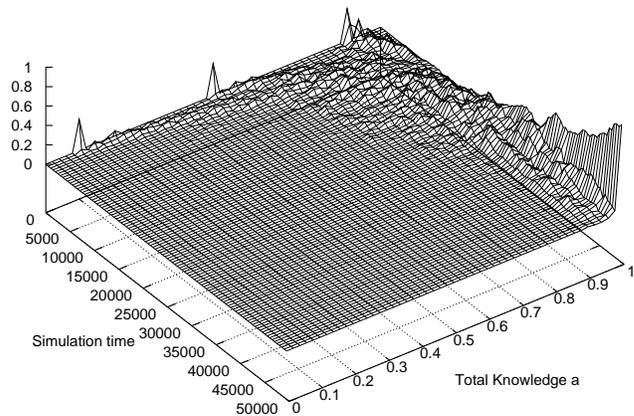}
\caption{\label{1E}Knowledge diffusion in a network of ``experts.''}
\end{figure}

We ran the experiment twice for populations with different personalities. In the first case (Fig.~\ref{1E}), the actors had high desire for reputation $\sigma=0.7$, low desire for knowledge $\kappa=0.2$, and very low desire for popularity $\pi=0.1$. We called these people ``experts.''

\begin{figure}[tb!]\centering
\epsfig{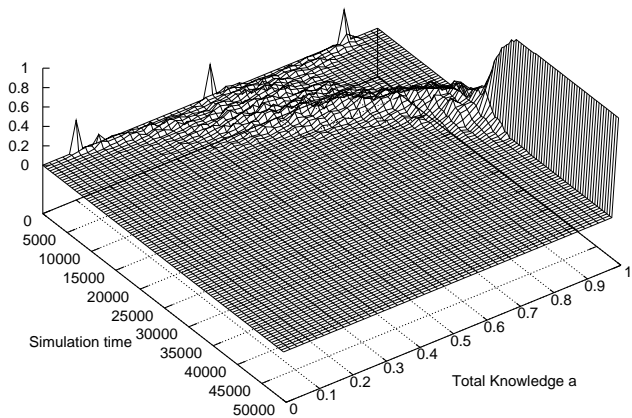}
\caption{\label{1T}Knowledge diffusion in a network of ``trolls.''}
\end{figure}

In the second experiment, the actors had low expectations for reputation and knowledge $\kappa=\sigma=0.1$ and strong popularity lust $\pi=0.8$ (Fig.~\ref{1T}). We called them ``trolls'' (this terminology was borrowed from~\cite{zinoviev2010a,zinoviev2010}).

As one can see from the figures, in both scenarios the distribution of knowledge rapidly converges to a situation where all network members become fully or almost fully knowledgeable ($a\approx 1$). In the network of ``trolls'' full convergence takes less time than in the network of ``experts'' (though, presumably, the quality of knowledge in the former network is substantially lower). Also, the ``expert'' network actually never fully converged to the full-knowledge state, always having some unsubstantial number of people doubting the diffused facts.

Another interesting difference between the two scenarios is the initial convergence of ``trolls'' to a medium-knowledge state, from where they proceed to the total-knowledge state. This means that in a ``troll'' community, actors absorb assertions without properly evaluating them. As a result, ``smarter'' actors become ``more stupid,'' and ``stupid'' actors become ``smarter,'' before the entire population becomes ``smarter'' as a whole.

The number of simultaneous receivers $N$ dramatically affected the convergence speed (more listeners result in faster convergence), but did not change the structure of the convergence.

\section{\label{conclusion}CONCLUSION AND FUTURE WORK}

In this paper we extended a previously developed game theoretical model of information dissemination in a star-like network with one sender and $N$ receivers. The model takes into account several personal traits of actors (the desire for knowledge, reputation, and popularity). The feedback mechanism is used to control the reputation of information senders and deter them from distributing unconfirmed rumors.

The model is mathematically represented by an $(N+1)$-player non-zero sum, non-cooperative game, where  the available actions are to forward an assertion or hold it indefinitely and to provide feedback on received assertions or not.

To improve and generalize our model, we propose the following future research directions:
\begin{itemize}
\item Consider the variability of $\kappa$, $\rho$, and $\pi$ for different actors in a network. 
\item Explore a connection between popularity $P$, reputation $R$ (or trust $T$), and knowledge $A$ and selected characteristics of real massive online social networks.
\item Develop a mechanism of preferential attachment in social networks based on mutual utility gain from information exchange.
\item Study the adaptation process that allows actors to change their $\kappa$, $\rho$, and $\pi$ parameters in order to maximize utility.
\end{itemize} 

\section*{Acknowledgment}
The authors are grateful to Prof. Honggang Zhang, Ali Yakamercan, and Taron Kondakindi of Suffolk University for their valuable input. This research has been supported in part by the College of Arts and Sciences, Suffolk University, through an undergraduate research assistantship grant.

\bibliographystyle{acm}
\bibliography{cs}
\end{document}